# A HYPOTHETICAL EFFECT OF THE MAXWELL-PROCA ELECTROMAGNETIC STRESSES ON GALAXY ROTATION CURVES

**D.D. Ryutov**[1*], **Dmitry Budker**[1,2,3], and **V.V. Flambaum**[1,4]

[1]*Helmholtz Institute, Johannes Gutenberg University, 55128 Mainz*
[2]*Physics Department, University of California, Berkeley 94720-7300, USA*
[3]*Nuclear Science Division, Lawrence Berkeley National Laboratory, Berkeley, California 94720, USA*
[4]*School of Physics, University of New South Wales, Sydney 2052, Australia*

**Abstract**

Maxwell-Proca electrodynamics corresponding to finite photon mass causes substantial change of the Maxwell stress tensor and, under certain circumstances, may cause electromagnetic stresses to act effectively as "negative pressure." The paper describes a model where this negative pressure imitates gravitational pull and may produce forces comparable to gravity and even become dominant. The effect is associated with random magnetic fields with correlation lengths exceeding the photon Compton wavelength. The stresses act predominantly on the interstellar gas and cause an additional force pulling the gas towards the center and towards the galactic plane. Stars do not experience any significant direct force but get involved in this process via a "recycling loop" where rapidly evolving massive stars are formed from the gas undergoing galactic rotation and then lose their mass back to the gas within a time shorter than roughly 1/6 of the rotation period. This makes their dynamics inseparable from that of the rotating gas. The lighter, slowly evolving stars, as soon as they are formed, lose connection to the gas and are confined within the galaxy only gravitationally. Peculiarities in the dynamics of these slowly evolving stars may serve as an experimental test of the presence and magnitude of the Maxwell-Proca stresses. In fact, observational data for sun-like stars in our galaxy appear to be incompatible with the assumption of the Maxwell-Proca stresses contributing as a level needed to explain the galactic rotation curve without appealing, for example, to possible galactic-evolution effects. It may be interesting to also explore possible broader cosmological implications of the negative-pressure model.

## 1. INTRODUCTION

The currently accepted upper bound for the photon mass (Tanabashi et al, 2018; Ryutov, 2007) is $m_{ph} < 2 \cdot 10^{-24} \, m_e \sim 10^{-18}$ eV, where $m_e$ is the electron mass. Therefore, the photon mass,

---
[*] Permanent address: VSP, L-440, Lawrence Livermore National Laboratory, Livermore, CA 94550, USA

even if finite, being so small, is normally ignored in the consideration of atomic and nuclear processes.

On the other hand, as noticed decades ago by Yamaguchi (1959), even a tiny photon mass may affect large-scale astrophysical phenomena, those occurring at length scales exceeding the photon Compton length

$$\lambdabar = \hbar / m_{ph}c ,\qquad(1)$$

where $\hbar$ is the reduced Planck constant and $c$ is the speed of light. For the currently accepted mass limit,

$$\lambdabar > 1 \text{ au}.\qquad(2)$$

The finiteness of the photon mass manifests itself at the scales exceeding $\lambdabar$ via the change of the Maxwell equations, which now explicitly contain vector and scalar potentials, and these potentials become measurable quantities. The corresponding equations, sometimes called Maxwell-Proca equations, are presented and discussed, for example, by Goldhaber & Nieto (1971) and Jackson (1975).

This change affects the dynamics of the interaction of a conducting medium with electromagnetic fields and leads to significant changes in magneto-hydrodynamic (MHD) phenomena, as discussed, for example, by Ryutov (1997). The limit (2) was obtained based on decades of direct observations of the Solar wind, most notably by the Voyager 1 and 2 missions (Ness & Burlaga, 2001). The limit (2) represents a firm *lower bound* on the photon Compton length.

Constraints based on directly observed and measured parameters (e.g., measurements of the spatial distribution of the magnetic fields of the Earth and other planetary bodies, or spacecraft measurements of the Solar-wind flow) are called "secure" in the review paper by Goldhaber and Nieto (2010).

One can attempt to obtain a stronger limit on $\lambdabar$ by considering phenomena at larger scales. Yamaguchi based his estimate on the presence of features of the scale of 0.03 pc in the Crab Nebula. Postulating that $\lambdabar$ must exceed this scale, he estimated that $\lambdabar > 0.03$ pc. Making specific assumptions on the magnitude of the interstellar magnetic field and its spatial structure, Adelberger et al (2007) suggested a lower bound of 1 kpc (see also Chibisov, 1976). However, in view of the fact that the assumptions used cannot be currently verified [see discussion in Sec. III of Goldhaber and Nieto (2010)], the values of the lower bounds obtained by these and some other approaches were called "speculative" by Goldhaber and Nieto (2010). The Particle Data Group in its annual reviews cites the "secure" limits.

As discussed below (for example, in Sec. 7.4), for our model to work, we need the photon Compton length in the range between a fraction of a pc to a few pc, i.e., not far from the range postulated by Yamaguchi. This, however, does not mean that we suggest a new limit for $\lambdabar$.

Interstellar medium is in a continuous exchange between the gaseous and stellar components of the galaxy [Ferrier (2001), Ryutov (2008)]. General considerations and observations show that the galactic gas is permeated by random magnetic fields [Crutcher (1999), Wielebinski & Beck (2005), Han (2006)]. Here our working hypothesis is that the correlation length of the random field exceeds the photon's Compton wavelength, which could be easily accommodated given the current lower limit on $\lambdabar$. Under these conditions, according to the Maxwell-Proca equations, the gas is effectively pulled towards the center of the galaxy and towards the equatorial plane, potentially contributing to the observed

"flat" rotation curves (Rubin, et al., 1980; Bosma, 1981). We emphasize that galaxies of different ages, sizes and angular momenta may differ in terms of the random magnetic field strengths and the relative role of Proca stresses and gravitational forces. Therefore, our mechanism may have quite different outcomes depending on the type and age of the galaxy. Generally, as the electromagnetic stresses affect mostly the gaseous component, the effects of these stresses should be particularly pronounced in galaxies with significant presence of gas.

The stars that are being formed from this gas "inherit" its velocity "at birth" and move initially together with the gas. If the stars' lifetimes (limited by their explosions as supernovae or by loss of most of their mass via intense stellar winds) are shorter by a factor of a few than their rotation periods around the galaxy center, the stellar material returns to the gaseous phase faster than one revolution occurs. In this scenario, the characteristic momentum-exchange time between the gas and the stars is short compared to the rotation period and the stars are strongly mechanically coupled with the gas: the forces acting on the gas are indirectly acting on the stars as well. The recycling time between the gas and the bright massive stars is indeed shorter than their period of rotation around the center of the galaxy (Ferriere, 2001; McGaugh & Blok, 1997; Kroupa, 2002; Chabrier, 2003; Bertulani, 2013).

Conversely, the stars with a lifetime greatly exceeding the rotation period are weakly coupled with the gas and do not experience a strong average pull towards the center. Such stars may freely travel to large radii and/or normally to the plane of the disk and even be lost. Although the Proca stresses can be quite significant and act in the "right" direction (towards the galaxy center), they cannot explain confinement of the slowly evolving stars, which do not experience the effect of the Proca stresses (even indirectly, via recycling) but are still confined on quasi-circular orbits. This shows that the Proca stresses, albeit potentially significant, have to be considered only in combination with other mechanisms of the centripetal pull (e.g., gravity). We address this issue in more detail in Sec. 6.

We emphasize that the presence of the Proca stresses in the galactic disks is just a hypothesis at present. However, given the current interest in the understanding of rotation curves of various galaxies (Rubin, et al., 1980; Bosma, 1981; Genzel et al; 2017), we feel it is useful to develop a more detailed and quantitative framework that would allow comparisons with the observations and other models. The current report is a first step in this direction.

The effects discussed in this paper are not those of dark matter: they are driven not by the *gravity force*, but by the peculiar *electrodynamic force* present in the Proca model and may co-exist with gravity-driven effects.

The paper is organized as follows. The next section describes the MHD equations for the Maxwell-Proca electrodynamics. Section 3 discusses a model of a random magnetic field with a correlation length significantly smaller than the global scale of the problem; it turns out that the forces produced by this field can be represented by gradients of some effective pressure that can become negative for a nonzero photon mass. The next step (Section 4) is the evaluation of the additional centripetal force acting on the galactic matter due to the negative pressure. It is shown that for the parameters of the Milky Way the effect can be significant. Section 5 discusses the electromagnetic forces directly acting (rather than via the recycling mechanism) on stars and molecular clouds; a conclusion is drawn that these forces are small and cannot affect the motion of these objects. Section 6 is concerned with the equilibrium in the direction normal to the galactic disk and the dynamics of slowly evolving stars. Here we evaluate the magnitude of the Proca stresses that would be compatible with the dynamics of slowly evolving stars for our Galaxy and find that the pull due to Proca stresses at the level of ~0.2 of the gravitational pull would be compatible with observations. Section 7 reiterates the main assumptions and addresses their plausibility. Finally, Section 8 contains summary and discussion. Supporting calculations are presented in Appendices A - D. Appendix E contains list of notations.

## 2. MHD EQUATIONS FOR THE MAXWELL-PROCA ELECTRODYNAMICS

We consider deeply nonrelativistic motion of conducting medium, as relevant to our problem. For such slow motion, one can use a quasi-static version of the Maxwell-Proca equations, where the displacement current and, correspondingly, the divergence of the current density are set to zero. For reference, the full set of the Maxwell-Proca equations is presented in Appendix A, where their reduction to the quasi-steady set is also discussed. In the quasi-static limit these equations read (throughout the paper we are using CGS units):

$$\nabla \times \boldsymbol{E} = -\frac{1}{c}\frac{\partial \boldsymbol{B}}{\partial t} \, , \qquad (3)$$

$$\nabla \times \boldsymbol{A} = \boldsymbol{B} \, , \qquad (4)$$

$$\nabla \times \boldsymbol{B} + \frac{\boldsymbol{A}}{\lambda^2} = \frac{4\pi}{c}\boldsymbol{j} \, , \qquad (5)$$

$$\nabla \cdot \boldsymbol{A} = 0 \, . \qquad (6)$$

The difference with the quasistatic version of the Maxwell equations is in the Ampere law [Eq. (5)], where an additional term appears on the left-hand side. The fields are created by slowly varying currents, with the characteristic time of the current variation greatly exceeding the transit time of light across the system. The current can be related to the fields by Ohm's law that does not change with respect to ordinary electrodynamics:

$$\boldsymbol{E} + \frac{\mathbf{v} \times \boldsymbol{B}}{c} = \eta \boldsymbol{j} \, , \qquad (7)$$

where $\eta$ is electrical resistivity of the medium, and $\mathbf{v}$ is velocity of the medium in the galactic frame.

In the problems that we consider in this paper the resistivity $\eta$ turns out to be low in the sense that the resistive dissipation time of the magnetic field is much greater than the relevant characteristic times. Accordingly, one can set $\eta$ to zero and use a model of the perfectly conducting medium, in which the electric field in the rest-frame of the medium is zero, i.e.,

$$\boldsymbol{E} + \frac{1}{c}\mathbf{v} \times \boldsymbol{B} = 0 \qquad (8)$$

Later, in Sec. 7.1 we provide more details on the accuracy of this approximation.

The electrodynamic equations are coupled with the equations of fluid dynamics

$$\frac{\partial \rho}{\partial t} + \nabla(\rho \mathbf{v}) = 0 ,\tag{9}$$

$$\rho \frac{d\mathbf{v}}{dt} = -\nabla p + \frac{1}{c} \mathbf{j} \times \mathbf{B} + \rho \mathbf{g} ,\tag{10}$$

where $\rho$ is the mass density, $p$ is the pressure, and $g$ is gravitational acceleration. The coupling occurs via the Lorentz force $\mathbf{j} \times \mathbf{B}/c$, which, using Eqs. (4, 5), can be represented also as

$$\frac{1}{c} \mathbf{j} \times \mathbf{B} = -\frac{\mathbf{B} \times \nabla \times \mathbf{B}}{4\pi} + \frac{\mathbf{A} \times \nabla \times \mathbf{A}}{4\pi \lambda^2} .\tag{11}$$

If necessary, one may also consider the energy equation for the gas:

$$\frac{\partial \varepsilon}{\partial t} + \mathbf{v} \cdot \nabla \varepsilon = -(\varepsilon + p) \nabla \cdot \mathbf{v} .\tag{12}$$

Here $\varepsilon$ is internal energy density of the matter, related to $p$ and $\rho$ by the equation of state

$$\varepsilon = \varepsilon(p, \rho) .\tag{13}$$

Equations (3-6), (9-10) and (12-13) form the required closed set of Maxwell-Proca MHD equations. Appendix B contains derivations of the energy integral and virial integral for this set of equations.

## 3. ELECTROMAGNETIC STRESSES IN A GAS CONTAINING SPATIALLY RANDOM MAGNETIC FIELDS

The magnetic field in the interstellar space is continuously perturbed by a variety of sources, in particular, supernova ejecta and intense stellar winds. It is, therefore, reasonable to assume that the magnetic field entrained by these motions becomes irregular, with some scale $l$ that may be related to the sources of the random flows. This scale is much smaller than the global scales of the problem which are the disk thickness $H$ and the distance to the galactic center $R$:

$$l << H < R.\tag{14}$$

The characteristic scale $l$ is probably in the range between a fraction of a parsec to a few parsecs (Ferriere, 2001; Crutcher, 1999; Wielebinski & Beck, 2005; Han, 2006). We refer to it as the correlation length due to the random nature of the magnetic field.

In assessing the galactic rotation curve and evolution of the disk thickness we are interested in the motion on the scales $H$ and $R$, which exceed by orders of magnitude the correlation length $l$. This motion is driven by the forces averaged over many correlation lengths. The corresponding averaging of the Lorentz force in the momentum equation [Eq. (11)] is performed in Appendix C. It turns out that the force can be presented in terms of the gradient of the effective magnetic pressure $P$:

$$\rho \frac{d\mathbf{v}}{dt} = -\nabla p - \nabla P + \rho \mathbf{g} .\tag{15}$$

Here $\mathbf{v}$ is now velocity averaged over small-scale motion; its dominant component is the rotation of the gas around the galaxy center. The magnetic pressure is (see Appendix C):

$$P = P_0 + P_1, \qquad (16)$$

where

$$P_0 = \frac{\langle B^2 \rangle}{24\pi}; \quad P_1 = -\frac{l^2}{\lambdabar^2} P_0. \qquad (17)$$

Note that, according to Eq. (1), $1/\lambdabar^2 \propto m_{ph}^2$.

One sees that the contribution to the pressure from the effect due to finite photon mass ($P_1$) can become significantly higher than the pressure arising from the Maxwell stresses ($P_0$) if the correlation length is larger than $\lambdabar$. As we show in the next section, the Maxwell stresses taken alone are too small to have any effect on the rotation curve, but the Proca addition can be significant for the sufficiently large ratio $l/\lambdabar$.

## 4. IMITATION OF THE GRAVITATIONAL PULL

As mentioned in the previous section, a finite photon mass may change the sign of the electromagnetic contribution to the momentum-flux tensor and, moreover, it may make the absolute value of the momentum-flux components large compared to the case of zero photon mass. This occurs if

$$l \gg \lambdabar. \qquad (18)$$

Let us now make numerical estimates to see what assumptions regarding the magnitude of various terms one has to make in order to have a noticeable effect on the rotation curve.

As a test case, we choose our galaxy and will focus on the properties of the galactic disk at the location of the sun, i.e., at a distance $R$ from the center of approximately 10 kpc; the rotation velocity at this distance will be taken as v=220 km/s. Assuming steady-state rotation, the centripetal acceleration at this distance is $a_{obs}=v^2/R$ (the subscript "obs" stands for "observed"). The average number density of protons in the interstellar medium will be taken as $n=0.3$ cm$^{-3}$; average temperature will be taken as $T=1$ eV and the magnitude of the random magnetic field $B$ as $2\cdot10^{-6}$G. Thus, we use the following reference parameters (cf. Ferriere, 2001):

$$n=0.3 \text{ cm}^{-3}; T=1 \text{ eV}; B\sim2\cdot10^{-6}\text{G}; v=220 \text{ km/s}; R=10 \text{ kpc}, 1/\Omega=R/v\sim5\cdot10^{7} \text{ yr.} \qquad (19)$$

One should not seek any accuracy in these values, this is just the first rough choice.

Note that the interstellar medium is actually not uniform and some parts of it, e.g., giant molecular clouds, may have much higher density. We discuss the effects of such "clumpiness" in the next section, but here we use the simplest model of uniformly spread highly conducting gas rotating around the galactic center.

The Proca correction to the net force acting on the gaseous disk is such that it imitates the gravity force, i.e., it pulls the matter in the "inward" direction. For the gaseous disk "inward" has two components: towards the rotation center and towards the mid-plane. The first may increase the rotation velocity, the second would make the disk thinner. We start from the radial pull. The model of the magnetic field in the rotating gas is that of random field-lines with some correlation length $l$ that is small compared tho the macroscopic dimensions of the disc (of which the smallest is the disc thickness $H$). In other words, we assume that $l \ll H$. The randomness is a natural assumption for the medium continuously "raked" by randomly distributed sources of energy and momentum, like SN explosions and outflows from the young stars.

There may exist also a regular magnetic field overlaying the random field. This regular field, due to its larger scale, has to be weaker than the random field by the ratio of the spatial scales (e.g. the radius of a fluxtube formed by the regular field to $l$), see Eqs. (16), (17). Experimental measurement of the large-scale magnetic field (on the background of the stronger random field) is a notoriously difficult task. From general considerations and numerical modeling (see in particular R. Pakmor et al, 2017), this field should be present on some level. It may play a role in formation of large-scale flux ropes and affect the formation of spiral arms. Some characteristic plasma equilibria in the presence of a global magnetic field were discussed by Ryutov, 1997. It would, however, be premature to try to develop a theory of the formation of global structures ahead of deeper characterization of galactic models in the presence of Proca stresses.

For a magnetic field with a randomly varying direction and a scale-length significantly shorter than the global scale, the resulting force is directed towards increasing $\langle A^2 \rangle$. One can surmise that the random magnetic fields and, accordingly, their random vector potential grow towards the center of the galaxy, with the length-scale $\sim R$. Using the results of Sec. 3, one then obtains the following estimate of the force $f$ acting in the radial direction on a unit volume of the gas:

$$f_R \sim \frac{|P_1|}{R} \sim \frac{\langle A \rangle^2}{24\pi \lambdabar^2 R} \sim \frac{\langle B \rangle^2 l^2}{24\pi \lambdabar^2 R} \ . \tag{20}$$

The resulting contribution to the centripetal acceleration is

$$a_R \sim \frac{f_R}{nm_p} \sim \frac{\langle B^2 \rangle l^2}{24\pi \lambdabar^2 R n m_p} \ . \tag{21}$$

The condition that this addition is comparable to the observed acceleration $a_{obs} = v^2/R$ then yields

$$\frac{l^2}{\lambdabar^2} \sim \frac{24\pi n m_p v^2}{\langle B^2 \rangle} \ . \tag{22}$$

For the reference parameters of the Eq. (19), we find that $l^2/\lambdabar^2 \sim 500$. Recalling that the energy density of the magnetic field in the case $l \gg \lambdabar$ is $W_1 = (l^2/\lambda^2)\langle B^2 \rangle / 8\pi$ [Eq. (C11)], we see from Eq. (22) that, in the state of rotational equilibrium provided by the Proca stresses, the energy density of the rotating gas is comparable to the magnetic energy:

$$W_1 \approx 3 m_p n v^2 \ . \tag{23}$$

Returning to Eq. (22'), we find that for the reference set of parameters (19), the ratio of $l/\lambdabar$ is large

$$l/\lambdabar \sim 22 \ . \tag{24}$$

Our model is based on the assumption that the correlation length $l$ of the random field is significantly shorter than the disk thickness

$$H \sim 0.5 \text{ kpc.} \tag{25}$$

Assuming, for example, that

$$l \sim 10 \text{ pc,} \tag{26}$$

one sees that condition (22) is satisfied for $\lambdabar \sim 0.4$ pc , i.e., for $\lambdabar$ well above the current lower bound on the photon Compton length, Eq. (2). In other words, our model for the set of parameters (19), (24), (26) is consistent with the current limit on the photon mass.

Note that Eq. (22) corresponds to the extreme assumption that the Proca stresses are responsible for the entire centripetal acceleration. For a more general case where both the Proca stresses and gravitation are present, one can characterize the contribution of Proca stresses to the centripetal acceleration by a parameter $\alpha < 1$, the ratio of the Proca contribution [$a_R$ of Eq. (21)] to the full (observed) centripetal acceleration $a_{obs}$,

$$\alpha = a_R / a_{obs} < 1. \tag{27}$$

For this more general case, one should add a factor $\alpha < 1$ to the right-hand sides of Eqs. (22) and (23).

## 5. FORCES ACTING ON COMPACT OBJECTS

The Maxwell-Proca stresses act on the gaseous component of the galaxy. In this section we evaluate forces acting on stars in order to evaluate possible contribution of these forces to the centripetal force. We consider two types of non-gravitational contributions: i) magnetic interaction of the magnetic field of the star with the large-scale fields, Sec. 5.1 and, ii) the drag force against the swirling gas (Sec. 5.2). We find that both forces are negligibly small and cannot affect the centripetal acceleration of the stars. This means that there exists a crucial difference between the dynamics of long-lived stars as compared to that of rapidly-evolving stars coupled to the gas by the recycling mechanism. For slowly evolving stars like our sun, the centripetal force has to be provided by gravity (including possible effects of dark matter).

In Sec. 5.3 we discuss forces acting on another class of relatively compact objects, giant molecular clouds, and conclude that the Proca force on them is also small, and their dynamics is largely determined by recycling.

5.1 Magnetic forces

Here we show that the force acting on a compact object, e.g. a star, with radius $r_s$ ("s" stands for a star) much smaller than the photon Compton length, $r_s \ll \lambdabar$, is small and does not have direct effect on the stellar dynamics in the galactic disk. We account only for the dipole component of the star's magnetic field as finer features produce only local fields and cannot contribute significantly to the force acting on the star as a whole.

The force with which the external field acts on a unit volume of stellar material is

$$\boldsymbol{f}_{ext} = \frac{[\boldsymbol{j}_s \times \boldsymbol{B}_{ext}]}{c}, \tag{28}$$

where $\boldsymbol{j}_s$ is a current carried by the stellar material and $\boldsymbol{B}_{ext}$ is the magnetic field created by external sources. We do not retain here the self-interaction of the stellar current and stellar magnetic field as it cannot create a net force and is balanced against the pressure and gravity forces in the star. Equation (28) is general in the sense that it holds for the Proca MHD; however, due to a small size of stars compared to $\lambdabar$, one can find them via classical electrodynamics; the Proca effects do not play any role in evaluating the force acting on a star.

The external field varies on the scale $l$ that is very large compared to the star's radius $r_s$. Then, the net force (the volumetric force integrated over the volume) is

$$\boldsymbol{F}_{ext} = -\nabla(\boldsymbol{\mu} \cdot \boldsymbol{B}_{ext}), \qquad (29)$$

where $\boldsymbol{\mu}$ is the magnetic dipole moment of the star. The direction of the force depends on the orientation of the dipole in a spatially varying external magnetic field. The magnitude of the force can be roughly estimated as

$$F_{ext} \sim \mu B_{ext} / l. \qquad (30)$$

We express the magnetic moment in terms of its magnetic field at the surface of the star, $B_s$. To do so, we note that the magnetic field produced by the magnetic dipole at its equatorial plane is $|\boldsymbol{B}_s| = \mu / r_s^3$ (Landau & Lifshitz, 1971), so that $\mu = r_s^3 B_s$, and

$$F_{ext} \sim B_s B_{ext} r_s^3 / l \qquad (31)$$

For this force to be comparable with the centripetal force (equal to $M_s v^2 / R$, where $M_s$ is the mass of the star), the condition

$$\frac{B_s r_s^3}{M_s} \sim \frac{l v^2}{B_{ext} R} \qquad (32)$$

has to be satisfied. However, for a star similar to the sun $B_s \sim 10\ G,\ r_s \sim 7.5 \cdot 10^{10}\ cm,\ M_s \sim 4 \cdot 10^{33}\ g$), and other parameters as in Eqs. (19) and (26), the left-hand-side of Eq. (32) is some 13 orders of magnitude smaller than the right-hand side, signifying that the direct magnetic force is negligible.

Consider now a force on a typical magnetar, a star of three solar masses and a radius of 20 km. For such a star, in order to have Eq. (32) satisfied, one has to assume the magnitude of the magnetic field $B_s \sim 2 \times 10^{32}$ G, many orders of magnitude higher than the highest magnetic field inferred for magnetars [see, for example, Harding&Lai (2006), Palmer et al (2005)]. Note also that, as the direction of the force is determined by the local magnetic field, not averaged over many randomly oriented parcels, the force will be randomly oriented. For all these reasons, we discard the role of the galactic magnetic field interaction with individual stars in the problem of the rotation curve.

5.2 Friction force

As the stellar objects do not experience the Proca force, they move only under the action of gravity and, generally speaking, have a velocity different from that of the gas. The relative motion (with a relative velocity $v_{rel}$) leads to a friction force. We now estimate the friction force to see if it can be strong enough to lead to entrainment effects.

For a star of a radius $r_s$ the friction force can be evaluated as $\pi r_s^2 m_p n v_{rel}^2$. Dividing by the mass of the star, we find its deceleration with respect to the ambient gas, $\dot{v}_{rel} \sim -\pi r_s^2 m_p n v_{rel}^2 / M_s$. This yields the following estimate for the slowing-down time: $\tau_{s-d} \sim v_{rel} / |\dot{v}_{rel}| \sim M_s / (\pi v_{rel} r_s^2 m_p n)$. For a numerical estimate, we take $v_{rel} \sim 0.2 v$ (see more details in Sec. 6); the other parameters are the same as in Sec. 5.1. This yields $\tau_{s-d} \sim 2 \cdot 10^{21} yr$ (!). In other words, the drag force is negligibly small and does not play any role in the dynamics of individual stars.

### 5.3 Forces acting on giant molecular clouds

Another group of compact galactic objects are giant molecular clouds (GMC; see, for example, Williams, et al, 2000; Cox, 2005). The mass of one such cloud may exceed that of $10^6$ solar masses. Their size is ~30 ps, and they are therefore much smaller in size than the thickness of the galactic disk. On the other hand, they are larger than the assumed scale of the random magnetic field structures of $l$~10 pc. They are thought to be the star-forming regions and are themselves formed from the explosions of massive supernovae. They are therefore involved in the recycling loop that allows to make the heavier and rapidly evolving stars to be involved in the gas rotation. As an aside, the clouds themselves probably contain quite strong and irregular magnetic fields [Ferriere (2001), Crutcher (1999)]. If these fields have correlation lengths exceeding $\lambda$, they could produce an inward pull, causing increase of the cloud density and of star-formation rate. This issue, however, goes beyond the scope of this paper.

As the size of a GMC is greater than the assumed correlation length of the external magnetic field, the Proca force acting on the GMC is determined by the momentum flux of this random field. Assume that the size of the cloud is some $r_c$ and the mass of the cloud is $M_C \sim \rho_C r_C^3$, where $\rho_C = m_p n_C$ is the cloud mass-density ($n_c$ is the number density of protons in the cloud; protons are assumed to dominate the cloud's particle content, although other elements and dust are also present). The force acting on the cloud via the momentum flux of the interstellar magnetic field can be found analogously to Eq. (20) as

$$F_C \sim P_1 r_C^2 \frac{r_C}{R} . \qquad (33)$$

The contribution of this force to the centripetal acceleration is

$$\Delta a_R = \frac{F_C}{M_C} \sim \frac{P_1}{\rho_C R} \sim a_R \frac{n}{n_C} . \qquad (34)$$

It is significantly smaller than $a_R$ as it contains a small factor $n/n_C$, typically ~1/300. In other words, the clouds experience a relatively small pull to the center of the galaxy and can follow the rotation of the interstellar gas only via fast recycling. The latter is plausible as a typical lifetime of these clouds is ~ 20 Myr [Ferrier, (2001), Cox, (2005)].

There is another mechanism that can affect the cloud dynamics: a friction force against the ambient gas (cf. Sec. 5.2). A cloud moving through the gas with a relative velocity $v_{rel}$ experiences a friction force ~ $\pi r_C^2 m_p n v_{rel}^2$. Dividing by the cloud mass, we find its deceleration (with respect to the ambient gas) $\dot{v}_{rel} \approx -(v_{rel}^2/r_C)(n/n_C)$. This yields the estimate for the characteristic slowing-down time: $\tau_{s-d} = v_{rel}/|\dot{v}_{rel}| \sim (r_C/v_{rel})(n_c/n)$. For a cloud diameter of 30 pc ($a_C \sim 30$ pc), $n/n_C$~1/300, and $v_{rel}$~100 km/s, this time is ~45 Myr, i.e. roughly twice longer than the typical cloud lifetime. A relatively short slowing-down time compared to that for stellar objects is related to much lower average particle density and much larger dimensions of the clouds.

# 6. RAPIDLY AND SLOWLY EVOLVING STARS, VERTICAL EQUILIBRIUM, AND GALACTIC HALO

As shown above, as soon as the density of a protostar becomes significantly higher than the density of the surrounding medium, so that the electromagnetic force becomes a few times smaller than the gravity force, the dynamics of the protostar (and, eventually, of the star) is controlled predominantly by gravity. For heavy, rapidly evolving stars [with masses exceeding ~ 3-4 solar masses (Bertulani, 2013), the stellar material returns to the interstellar gas within a time shorter than a fraction of one revolution time: within this time the star explodes as a supernova or loses its mass to the stellar wind and thereby returns its material to the interstellar gas before the star's velocity significantly departs from that of the rotating gas (fast recycling).

Consider now light, slowly evolving stars, for which recycling does not take place and which therefore move on orbits determined by the gravity force alone. If the Proca force were dominant, then the slowly evolving stars, upon their condensation from the gas, would not experience any significant centripetal force and would leave the galaxy ballistically, as a stone launched from a sling. This is obviously not the case for our galaxy. We therefore conclude that in the case of our galaxy the Proca forces, if present, can constitute only a fraction of the total centripetal force. In such a case, there will be some difference of the rotation speed of the gaseous component and long-lived stars at every radius $R$. The presence of such difference, even if modest, may serve as an experimental "detector" of the magnitude of the Proca stresses. The sign of this difference corresponds to faster rotation of the gaseous component, whereas the magnitude depends on the distribution of the gravitating mass (including, possibly, dark matter).

To be specific, we consider the situation where the (Newtonian) gravity is created by a central mass situated well within the radius $R$. The situation where the gravity is produced by a diffuse halo (made possibly of dark matter) that encompasses the star's trajectory will be considered elsewhere. In the case of a dominant central mass, a light star produced from the rotating gas proceeds along an elliptical trajectory (Fig. 1).

Let $R_0$ be some reference radius, where the rotation velocity of the gas is $v_0$. This means that its centripetal acceleration is $v_0^2 / R_0$. A fraction $\alpha$ of this acceleration is produced by the Proca stresses, and $(1-\alpha)$ is produced by gravity. Then at other radii, the centripetal acceleration $a$ will be

$$a = \frac{v_0^2}{R_0}\left[(1-\alpha)\left(\frac{R_0}{R}\right)^2 + \alpha\left(\frac{R_0}{R}\right)\right]. \qquad (35)$$

Here we assumed that the Proca term, if taken alone, $\alpha = 1$, would produce a flat rotation curve, Eq. (21), but we are going to consider a situation, where the first term in the square bracket is dominant, $\alpha < 0.3$.

When a light (and long-lived) star is promptly formed from the gas at point "0" in Fig 1, it has a velocity $v_0$, but does not any longer experience the Proca force. It continues under the action of the central gravitating mass [the first term in Eq. (35)]. Its orbit becomes an ellipse, with a maximum distance from the center $R_1 > R_0$. We imply that $\alpha<0.5$, as otherwise the motion would become unconfined, with a star flying away to infinity [see Eq. (36) below]. The crossing time of the distance of $R$~10 kpc for a freely moving star with velocity of 220 km/s, Eq. (19), would be only 50 Myr, and the Galaxy would have lost the solar-type stars (including the sun itself) long ago. This is not the case, and this is the reason why we are forced to assume that $\alpha$ is small ($\alpha < 0.3$).

The maximum distance $R_1$ from the center would be determined by conservation of energy (with the last term in Eq. (35) turned off) and angular momentum. In this way one finds that

$$\frac{R_1}{R_0} = \frac{1}{1-2\alpha}; \quad v_1 = v_0(1-2\alpha), \qquad (36)$$

where $v_1$ is the velocity at $R_1$.

Velocity of the gas performing circular motion along the circle with $R=R_1$ under the action of both gravitational and Proca terms is, according to Eq. (35), equal to:

$$v_{1g} = \sqrt{aR_1} = v_0\sqrt{(1-\alpha)(R_0/R_1)+\alpha} = v_0\sqrt{(1-\alpha)^2+\alpha^2}. \qquad (37)$$

A star quickly formed at point "1" from the rotating gas would have velocity $v_{1g}$. The stars formed at intermediate azimuthal angles would have intermediate azimuthal as well as some radial velocities. Without going into exercise of deriving the velocity distribution function of light stars at point 1, we approximately evaluate an average azimuthal velocity of the slowly evolving stars at a radius $R_1$ as

$$\bar{v} = (v_{1g} + v_1)/2 \approx v_0\left(1 - \frac{3\alpha}{2}\right). \qquad (38)$$

The deviation of the azimuthal velocities from average will, accordingly, lie in the interval $\pm \alpha v_0/2$. Indeed, $v_{1g} - \bar{v} = \alpha v_0/2$, and $v_1 - \bar{v} = -\alpha v_0/2$. The standard deviation will be less than $\alpha v_0/2$, because all the intermediate velocities are also present. Without getting into more formal analysis, we take this factor into account by introducing another multiplier of ½ in the standard deviation:

$$\sqrt{\langle(\Delta v)^2\rangle} \approx \alpha v_0/4. \qquad (39)$$

For rapidly recycled heavy stars with a lifetime $\tau \ll 1/\Omega$, the velocity spread will be much smaller.

Equation (39) can be used to constrain the allowable level of the Proca stresses that would not contradict the observed velocity dispersion (Binney & Merrifield, 1998). A 10-20 percent contribution ($\alpha$=0.1-0.2) does not lead to any contradictions. To be consistent with the observations, the contribution of the Proca stresses in our galaxy has to be small. Note also, that these stresses do not necessarily have to vary as $1/R$, this leading potentially to increase of the allowable magnitude of these forces. Note also that including the evolutionary effects may bring new factors and amplify the role of the Proca stresses. At this point it would be fair to say that in the case of our galaxy at its present state they cannot play a dominant role in its dynamical equilibria, staying at the level of less than 30 percent of the gravitational forces on the global scales. When considering the possible role of Proca stresses in more local phenomena, e.g., the structure of the galactic arms, one should remember that these stresses act directly only on the gaseous component of matter.

Average (over non-uniformities of the scale $l \ll H$) electromagnetic force acting in the "vertical" direction (normal to the disc plane) vary in this direction on the scale $H$ and are for this reason stronger by a factor of $R/H$ than the radial forces (causing the rotation of the gas and of rapidly evolving stars tightly bound to it). In a strictly one-dimensional picture this could lead to too small disk thickness (smaller than the commonly accepted $H \sim 0.05\,R$): a point made to the authors by C. McKee). Under a strong compression of the gas in the vertical direction, the Proca stresses should become anisotropic, as discussed in Appendix D, with the vertical pressure being smaller than that that acting in the disk plane.

Another explanation could be in dynamical nature of the vertical structure, with a significant waviness that, when averaged over several wavelengths, mimics the finite thickness. The source of this waviness may be the vertical motion of rapidly evolving molecular clouds and heavy stars that continuously stir the disk. Again, the vertical Proca force is essential only for the rapidly-evolving objects that maintain close connection to the motion of the interstellar gas. Slowly evolving stars, as soon as they are formed in a moving gaseous cloud, do not experience rapid recycling and are confined only by relatively weak gravity. They may therefore leave the disk and show up in the galactic halo.

A star, formed from gas that experiences intense vertical convection, acquires the gas velocity "at birth". It then continues subjected only to a weak gravitational pull. These gas-star transitions can add interesting and potentially observable features to the effect of galactic fountains (Shapiro & Field, 1976; Wakker & van Woerden, 1997).

Systematic analysis of these effects goes well beyond the scope of this paper. We only note that the analysis of the stellar motion as sketched in this section could potentially serve for distinguishing the relative effect of gravity and Proca stresses in galaxies.

## 7. DISCUSSION OF OUR MAIN ASSUMPTIONS

7.1 Slowness of Resistive Dissipation of the Random Magnetic Field

The effects considered in our paper are based on the interaction of magnetic field with the plasma current flowing in the interstellar medium. It is thus important to check whether the current-carrying capabilities of this tenuous medium are sufficient for sustaining the required currents. In this section we address this issue and show that the current limitations do not constrain our model in any significant way.

For magnetic field with length-scale $l > \lambdabar$ the current density determined by Eq. (5) is approximately

$$j = \frac{c\mathbf{A}}{4\pi \lambdabar^2}. \qquad (40)$$

We now perform some order-of-magnitude estimates. From Eq. (4) we can relate vector potential $A$ and magnetic field $B$: $A \sim Bl$. Equation (40) then shows that the current density, for a given $B$ and $l > \lambdabar$ is

$$j \sim \frac{cBl}{4\pi \lambdabar^2}. \qquad (41)$$

Using Eq. (7), evaluating $E$ from Eq. (3) as $E \sim Bl/c\tau_{res}$, where $\tau_{res}$ is the resistive dissipation time, and expressing the current via the Ohm's law, $E \sim \eta j$, we find the relation between current and magnetic field: $j \sim Bl/\eta c \tau_{res}$. Finally, substituting $j$ into Eq. (41), we get:

$$\tau_{res} \sim \frac{4\pi}{c^2 \eta} \lambdabar^2 . \tag{42}$$

For a more formal discussion see Ryutov (1997).

Consider as an example an ionized interstellar gas of a temperature $T=1$ eV (Ferriere, 2001, Table 1). Its resistivity in the CGS units is

$$\eta(\text{CGS}) \sim 10^{-13} [T(\text{eV})]^{-3/2} \sim 10^{-13} \text{s} \tag{43}$$

(Book, 1987). Taking $\lambdabar \sim 0.4$ pc [see Eqs. (24) and (26)], we find that for our reference case $\tau_{res} \sim 10^{29} \text{s} \approx 3 \times 10^{15}$ Myr i.e., many orders of magnitude longer than the age of the universe.

One can invert the problem and evaluate the resistivity $\eta^*$ at which the resistive dissipation time becomes equal to the typical recycling time $\sim 20$ Myr used in Sec. 5 and 6. Substituting $\tau_{res} \sim 20$ Myr and $\lambdabar \sim 0.4$ pc into Eq. (43), we find $\eta^* \sim 30$ s, i.e. 14 orders of magnitude higher than the resistivity of 1 eV plasma. The margin in this example is enormous.

Another way to look at the current density requirements is to evaluate the relative velocity $u$ of the electrons and ions, as at high values of $u/v_{Te}$, where $v_{Te}$ is electron thermal velocity, so called "anomalous resistance" may become significant, see, for example, Papadopoulos (1977). The current density is determined by Eq. (41) and, on the other hand, by $j = enu$. One finds that

$$\frac{u}{v_{Te}} \sim \frac{cBl}{4\pi env_{Te} \lambdabar^2} , \tag{44}$$

For our reference set of parameters (19), (24) and (26) we find that $u/v_{Te} \sim 10^{-11}$, many orders of magnitude less than the critical value of $u/v_{Te} \sim \sqrt{m_e/m_p} \sim 1/40$, required for the onset of plasma microinstabilities and anomalous resistance (Papadopoulos, 1977).

This extreme smallness of both $\eta/\eta^*$ and $u/v_{Te}$ accommodates possible increases of resistivity related to lower temperatures and lower ionization degrees, which are conditions that may be encountered in the lower-temperature areas atomic and molecular hydrogen in the H and $H_2$ regions (Table 1 in Ferriere, 2001). This allows one to use the perfect-conductivity model throughout the interstellar space.

The electrical conductivity in the dense molecular clouds may be determined by other effects, in particular, by the presence of dust and negative ions. Still, the resistivity inside the clouds is thought to be sufficiently low to prevent the magnetic field from being dissipated within their characteristic lifetime.

## 7.2 Force-Free Structure of Tangled Magnetic Field

Another issue is the plasma equilibrium within each "cell" of the scale $l$ where the magnetic field has a regular character. If one uses a naïve estimate for the force acting on a unit volume of the plasma within this cell,

$$f \sim \frac{jB}{c} , \tag{45}$$

one finds that the plasma inside each cell would start moving with velocity

$v \sim \sqrt{fl/mn_p}$. Using Eq. (42), we find:

$$v \sim \frac{B}{\sqrt{4\pi m_p n}} \times \frac{l}{\lambdabar} \equiv v_A \times \frac{l}{\lambdabar}, \quad (46)$$

where $v_A$ is the Alfven velocity. For the reference set of parameters (19), $v_A \sim 10$ km/s, comparable to the plasma sound velocity in $T \sim 1$ eV plasma. However, in Eq. (46) there is an additional large multiplier $l/\lambdabar \sim 22$. This seems to indicate that plasma in each parcel of a coherent magnetic field would be involved in a violent supra-thermal motion which would cause a further rapid entanglement of the magnetic field and its eventual dissipation. The time-scale of this process would be $l/v \sim 4 \cdot 10^4$ years, orders of magnitude shorter than the times $\sim 4 \cdot 10^7$ years that are of significance in the problem under consideration.

However, plasma is known to form force-free structures. The formation of such structures is, in some sense, natural, as the non-force-free components would be quickly dissipated in intense plasma flows generated by them, and only the force-free components would survive (Taylor, 1986). In "standard" (non-Proca) magnetohydrodynamics (MHD) force-free fields are encountered in plasma confinement devices and in Solar atmosphere, among other systems [Chandrasekhar & Woltjer (1958); Parker (1979); Wiegelmann & Sakurai (2012), Bellan (2012)]. It is conjectured that the force-free states may exist in the form of "magnetostatic turbulence" – a random magnetic field that is in a force-free equilibrium and thereby does not drive rapid, as in Eq. (46) plasma flows (Ryutov & Remington, 2007). This system still produces the same average momentum fluxes on the global scale as those estimated in Secs. 4, 5.

Virial arguments show that global equilibrium of such systems requires non-magnetic forces to hold the system together. In particular, in the magnetic confinement devices the magnetic field momentum flux from the force-free zone is countered by the stresses in the magnetic coils. In the case of interest for astrophysics, this can be a gravitational force holding together a plasma cloud filled with tangled, force-free magnetic field.

For the situation of $l \gg \lambdabar$ considered in our paper, the magnetic field tends to "pull" the matter inward. It has to be balanced by the centrifugal force in the radial direction and dynamic pressure of the matter in the normal (to the disk) direction.

For the magnetic field to be force-free, the condition of collinearity of plasma current and the magnetic field, $\boldsymbol{j} \parallel \boldsymbol{B}$, has to be satisfied. In the standard electrodynamics ($l \ll \lambdabar$) one has

$$\boldsymbol{j} = \frac{c}{4\pi} \boldsymbol{B}, \quad (47)$$

and the force-free equilibria are therefore defined by an equation

$$\boldsymbol{B} \times \nabla \times \boldsymbol{B} = 0. \quad (48)$$

In the case of interest for us, $l \gg \lambdabar$, Eqs. (4) and (5) show that one can neglect $\nabla \times \boldsymbol{B}$ in the left-hand side of Eq. (5), thereby leading to equation

$$\boldsymbol{j} = \frac{c}{4\pi \lambdabar^2} \boldsymbol{A}, \quad (49)$$

and, as $\boldsymbol{B} = \nabla \times \boldsymbol{A}$, the force-free equilibrium is defined by an equation

$$\boldsymbol{A} \times \nabla \times \boldsymbol{A} = 0. \quad (50)$$

Therefore, the "standard" theory of the force-free fields [Chandrasekhar & Woltjer (1958); Taylor, (1974, 1986)] can be applied to our system simply by a substitution $\mathbf{B} \rightarrow \mathbf{A}$. So, if the states of magnetostatic turbulence exist in the ordinary MHD, they should exist in our system as well.

An assumption of an almost force-free state of the tangled magnetic field is a key (and not rigorously proven yet) assumption of our analysis and should at present be considered as such [although numerous other examples of the force-free states summarized in Chandrasekhar & Woltjer (1958); Parker (1979); Wiegelmann & Sakurai (2012), Bellan (2012) do, indeed, exist].

## 7.3 Insignificance of the Contribution of the Electromagnetic Field Energy to Gravity

The magnetic field (whose energy density in the case $l \gg \lambdabar$ is equal to $W_1$) provides some contribution $\Delta\rho_1 = W_1/c^2$ to the mass density of the disk. Equation (23) shows that this energy density is approximately equal to the rotational energy density of the gas, $W_1 \sim m_p n \mathrm{v}^2$. Therefore,

$$\frac{\Delta\rho_1}{m_p n} \sim \frac{\mathrm{v}^2}{c^2} \ll 1. \tag{51}$$

In other words, the additional mass density is negligibly small and does not contribute to the gravity: our effect is caused by the negative pressure of the tangled magnetic field in the $l \gg \lambdabar$ case, not by an additional gravity force.

According to Eq. (23), magnetic-field energy is comparable to the kinetic energy of the rotating gas. This is small by many orders of magnitude compared to the energy released by energetic events during one recycling period (Ferriere, 2001). Correspondingly, the magnetic energy has no significant effect on the energy balance of the system.

## 7.4 Constraints on the Photon Mass

We use the photon's reduced Compton length $\lambdabar$ as a measure of its mass [see Eq. (1)]. Our approach in this paper relies on the assumption of separation of scales: the correlation length $l$ has to be much smaller than the galactic-disk thickness $H$ and much larger than $\lambdabar$. This imposes constraints on the range of photon masses that would be consistent with our model. Indeed, $\lambdabar$ has to be smaller than $l$, by roughly a factor of 25 (Sec. 4), to make the forces related to modified electrodynamics significant for typical values of galactic magnetic fields. On the other hand, too small $\lambdabar$ for a given $l$ would make the magnetic forces too strong and again incompatible with the observations. A plausible assumption is $\lambdabar \sim l/25$. But the length $l$ is also constrained: as mentioned, it should be by a factor of ~10 less than $H$, i.e., $l_{upper} \sim 0.1H$. The natural lower bound for $l$ is hard to evaluate due to the uncertainties associated with the magnetic-dynamo effects (Zeldovich, et al, 1990) perhaps, $l_{lower} \sim 1$ pc is a reasonable value. With this assumption, the range of $\lambdabar$ compatible with our model is between $0.04 l_{lower} \sim 0.04$ pc and $0.04 l_{upper} \sim 4 \cdot 10^{-3} H \sim 2$ pc. This corresponds to photon masses that are some 4-6 orders of magnitude lower than the current "safe" upper bound (Tanabashi et al, 2018).

# 8. SUMMARY AND DISCUSSION

This paper considers a possible role of the photon mass in the rotational dynamics of galaxies. The basic mechanism is related to the additional centripetal force provided by the magnetic field stresses enhanced by the Proca mechanism. The effect is, therefore, not an effect of additional gravity force. It may act in parallel with the effect of the dark matter and, in some range of parameters, may completely eliminate the need for the dark matter for the explanation of the rotation curve (although this is not the case for our galaxy).

The model discussed in this paper relates the effect of a possibly finite photon mass to the change in the magnetic stresses acting on the conducting galactic gas. The dominant magnetic field is assumed to be random. If the spatial scale $l$ of this random field is greater than the photon Compton length $\lambdabar$ [Eq. (1)], the stresses (that we call "Proca stresses") can be described as negative pressure pulling the gas towards stronger magnetic field. In the galactic environment, this pull has two components: the radial one, pulling the gas to the galactic center, and "vertical" one, pulling the gas to the galactic plane. The radial pull becomes comparable to the gravitational pull if condition (22) is fulfilled. If so, the pull produced by the Proca stresses may partially substitute the gravitational pull and thereby affect the rotation curve. The main unknown parameter here is the ratio $l/\lambdabar$. For our galaxy, a significant effect at the location of the Solar system requires $l/\lambdabar \sim 20$. In this case, the energy density of the magnetic field [Eqs. (22) and (C.11)] is on the order of rotation energy density of the gas, $m_p n v^2$.

This additional Proca pull is negligible for compact objects, like protostars and stars. Our mechanism affects these objects indirectly: as a compact object is formed from the rotating gas, it acquires the rotational velocity of the gas at the time of formation of the compact object. The formation time of a protostar is typically less than a few million years [McKee & Ostriker (2007)] and is negligible compared to $2\pi/\Omega \sim 10^8$ years. In other words, a compact object at its birth moves with the same velocity as the gas. Heavy stars, whose evolutionary time is shorter than $2\pi/\Omega \sim 10^8$ years, lose almost all their mass to the stellar winds and/or supernova explosions within the time shorter than $1/\Omega$, returning their material to the interstellar gas. This "recycling" process provides strong coupling of the heavy stars (which make the dominant contribution to the luminosity of the galaxies) with the gas that is pulled towards the rotation center by the Proca magnetic forces.

The situation is entirely different for light stars (with masses below a few solar masses). For them, the evolutionary time is much longer than one full rotation period $2\pi/\Omega$. In the absence of gravitational force, these stars would leave galaxy ballistically. As in our galaxy these lighter stars are rotating not significantly differently from the gas and heavier stars, this imposes an upper bound on the possible contribution of the Proca stresses, as discussed in Section 6. The effect of the Proca stresses is then a somewhat slower rotation of light stars vs. the gas and heavier stars at the same distance from the center.

The Proca stresses provide also a vertical force that confines the gas in the vicinity of the equatorial plane. The recycling mechanism mentioned above also confines rapidly evolving stars near the equatorial plane. As before, this effect is small for slowly evolving stars, some of which, depending on the initial conditions, escape from the equatorial plane and gradually create a stellar population of the galactic halo.

As emphasized in the Introduction, the Proca stresses affect only the dynamics of the gaseous component of a galaxy and, indirectly, via the recycling process, the dynamics of short-lived bright stars. For distant galaxies, those stars are indeed the sources of information on the rotation curves. In the case of our galaxy, there exists also detailed information on the dynamics of lighter, long-lived stars (e.g., Binney & Merrifield, 1998). In this case one can attempt to distinguish between the contributions of dark matter and Proca stresses from the presence of a relatively large peculiar velocities of these stars as discussed in Sec. 6.

Although all these effects are related to the finiteness of the photon mass, they are nogravitational in their essence: the mass of the Proca field is negligible with respect to the mass of the baryonic matter and does not make any significant contribution to the gravity force.

The presence and the structure of the random magnetic fields can be different for various types of galaxies, depending, in particular, on the age of the galaxy and relative amount of gas. The scale $l$ of magnetic non-uniformities may be determined by the frequency and spatial distribution of the "energetic events" like supernova explosions and formation of bright heavy stars with strong stellar winds. For a given photon mass (given $\lambdabar$) and galactic gas density, there exists a range of correlation lengths $l$ and field strengths where Proca stresses can be significant

As emphasized above, the model considered in this report is not that of dark matter; it may, however, account for the conjectured dark-matter effects in astrophysical settings even beyond galactic rotation curves. It may also be tempting to explore the influence of the "negative pressure" typical for the Proca electrodynamics on the cosmological-scale phenomena with a possible impact on the dark-energy scenarios. To have a significant effect, one would need the presence of a tangled magnetic field, with a correlation length exceeding the photon Compton length, filling the entire space. The field would have to be force-free to survive on the cosmological time-scales.

# Acknowledgements

The authors acknowledge critical feedback from C. McKee and S. Rajendran and discussions with M. Kamionkowski, D. Jackson Kimball, M. Pospelov and M. Zolotorev. D.D.R. acknowledges the hospitality of the Helmholtz Institute Mainz. D.B. acknowledges the support by the DFG Reinhart Koselleck project, the ERC (Dark-OST Advanced Project), and the Simons and the Heising-Simons Foundations. V.V.F. has been supported by the Australian Research Council (ARC) and the JGU Gutenberg Research Fellowship. D.B. and V.V.F. performed part of the work at the Aspen Center for Physics, which is supported by National Science Foundation grant PHY-1607611.

# APPENDIX A. QUASI-STEADY-STATE MAXWELL-PROCA EQUATIONS

A set of Maxwell-Proca equations reads [see Jackson, 1975; Goldhaber & Nieto, 1971]:

$$\nabla \times \boldsymbol{E} = -\frac{1}{c}\frac{\partial \boldsymbol{B}}{\partial t} \ ; \tag{A1}$$

$$\boldsymbol{B} = \nabla \times \boldsymbol{A} \ ; \tag{A2}$$

$$\nabla \times \boldsymbol{B} + \frac{\boldsymbol{A}}{\lambdabar^2} = \frac{4\pi}{c}\boldsymbol{j} + \frac{1}{c}\frac{\partial \boldsymbol{E}}{\partial t} \ ; \tag{A3}$$

$$\nabla \cdot \boldsymbol{E} + \frac{\varphi}{\lambdabar^2} = 4\pi\rho \ . \tag{A4}$$

Here $\boldsymbol{A}$ and $\varphi$ are the vector and scalar potentials, and $\rho$ and $\boldsymbol{j}$ are the space charge and the current density satisfying the standard continuity equation

$$\frac{\partial \rho}{\partial t} + \nabla \cdot \boldsymbol{j} = 0 \ . \tag{A5}$$

Note that we use here the same notation for the charge density as we have used in the main text for the mass density of the matter. This, however, should not cause confusion because the charge density does not appear in any equations of the main text.

The electric field is related to $\boldsymbol{A}$ and $\varphi$ by the following equation:

$$\boldsymbol{E} + \frac{1}{c}\frac{\partial \boldsymbol{A}}{\partial t} = -\nabla \varphi \ . \tag{A6}$$

The potentials satisfy the Lorentz gauge

$$\nabla \cdot \boldsymbol{A} + \frac{1}{c}\frac{\partial \varphi}{\partial t} = 0 \ . \tag{A7}$$

This (Lorentz) gauge becomes now the only possible one: Equation (A7) is a direct consequence of Eqs. (A3), (A4) and (A5). The scalar and vector potentials become measurable quantities.

Consider a system where the temporal variation of the "sources" (currents and charges) is slow in the sense that the characteristic time of this variation, $\hat{T}$, is long compared to the light-propagation time over the spatial scale of the system, $L$:

$$\hat{T} \gg L/c \tag{A8}$$

(we use the hat sign to distinguish the characteristic time from the temperature). Then, according to Eq. (A1), one has the following order-of-magnitude relation:

$$E \sim (L/c\hat{T})B \ . \tag{A9}$$

If one associates with $L$ and $\hat{T}$ some velocity $v = L/\hat{T}$ that characterizes the motions in the system, then one can say that $E$ is first-order in $v/c$. Substituting the estimate (A9) into Eq. (A3), one sees that the last term in the right-hand side is small compared to the first term in the left-hand side by a factor of $(v/c)^2$ and constitutes a correction of the second order. Neglecting this second-order term, one obtains the quasistatic version of Eq. (A3):

$$\nabla \times \boldsymbol{B} + \frac{\boldsymbol{A}}{\lambdabar^2} = \frac{4\pi}{c}\boldsymbol{j} \ . \tag{A10}$$

Likewise, estimating *A* from Eq. (A2) as *BL* and using Eq. (A9), we find from Eq. (A6) an order-of-magnitude estimate $\varphi \sim (L^2/c\hat{T})B$. Then, comparing the first and the second terms in Eq. (A7), we find that the second term is second-order small and can be dropped, so that Eq. (A7) is reduced to

$$\nabla \cdot \boldsymbol{A} = 0 . \tag{A11}$$

Taking the divergence of Eq. (A.10), one finds that, to the same (first) order, the charge-continuity equation is reduced to

$$\nabla \cdot \boldsymbol{j} = 0 . \tag{A12}$$

Equations (A1) and (A2) are the same in the quasi-steady-state approximations. Thereby we obtain a set of equations describing quasi-steady-state processes in the Maxwell-Proca electrodynamics. The set consists of Eqs. (A1) and (A2) and Eqs. (A10) and (A11). This set is reproduced in a compact form in Sec. 2 of the main text, Eqs. (3)-(6). Note that in the limit of a massless photon, $\lambdabar \to \infty$, Eqs. (3)-(6) become Maxwell equations for the quasi-steady-state processes.

Note that for quasi-static fields, the vector potential and the magnetic field exponentially decay $A \propto \exp(-r/\lambdabar)$ at large distance from the localized current system generating them [see Eqs. (A2) and (A10)]. This fact was used to constrain the photon mass by observing the magnetic field of the compact objects, Earth (Schrödinger, 1943) and Jupiter (Davis et al., 1975).

# APPENDIX B. ENERGY EQUATIONS AND VIRIAL INTEGRALS FOR THE MAXWELL-PROCA MAGNETO-HYDRODYNAMICS

In this Appendix we present calculations describing the energy integral for the Maxwell-Proca magneto-hydrodynamics of Sec. 2, discuss the virial theorem, and present averaging of the force terms over the scales including many correlation lengths *l*. It turns out that it is convenient for these calculations to combine Eqs. (3) and (9) and eliminate the electric field. The result is:

$$\frac{\partial \boldsymbol{B}}{\partial t} = \nabla \times [\boldsymbol{v} \times \boldsymbol{B}] . \tag{B1}$$

In the MHD parlance this equation is called the "line-tying equation" as it shows how the magnetic field is entrained by the moving perfectly conducting fluid.

When assessing the energy equation, we focus on the fluid and field energy and ignore the gravitational term. The corresponding details can be found in Landau & Lifshitz (1987). To derive the energy equation, we take a scalar product of Eq. (10) with the fluid velocity **v**. We will separately assess three terms: the left-hand side, the pressure term, and the Lorentz force term. For the first one we have:

$$\rho \boldsymbol{v} \cdot \frac{\partial \boldsymbol{v}}{\partial t} + \rho \boldsymbol{v} \cdot (\boldsymbol{v} \cdot \nabla) \boldsymbol{v} \equiv \frac{\partial}{\partial t}\left(\frac{\rho v^2}{2}\right) + \nabla \cdot \left(\boldsymbol{v} \frac{\rho v^2}{2}\right) - \frac{v^2}{2}\left(\frac{\partial \rho}{\partial t} + \nabla \cdot \rho \boldsymbol{v}\right) =$$

$$\nabla \cdot \left(\boldsymbol{v} \frac{\rho v^2}{2}\right) + \frac{\partial}{\partial t}\left(\frac{\rho v^2}{2}\right). \tag{B2}$$

At the last step in this transformation we used the continuity equation (9).

The pressure term is transformed as follows:
$$\mathbf{v}\cdot\nabla p = \nabla\cdot(\mathbf{v}p) - p\nabla\cdot\mathbf{v} \quad . \tag{B3}$$
Now using Eq. (13), one can reduce it to [cf. Landau & Lifshitz (1987)]:
$$\mathbf{v}\cdot\nabla p = -\nabla\cdot(\mathbf{v}w) - \frac{\partial\varepsilon}{\partial t}, \tag{B4}$$
where $w$ is enthalpy per unit volume. For polytropic gas one has $w = \gamma p/(\gamma-1)$, $\varepsilon = p/(\gamma-1)$, where $\gamma$ is the adiabatic index.

To transform the Lorentz term, we express $\mathbf{j}$ in terms of $\mathbf{A}$ and $\mathbf{B}$ via Eq. (5) and use Eq. (11):
$$\frac{\mathbf{j}\times\mathbf{B}}{c} = -\frac{\mathbf{B}\times\nabla\times\mathbf{B}}{4\pi} + \frac{\mathbf{A}\times\nabla\times\mathbf{A}}{4\pi\lambdabar^2} \quad . \tag{B5}$$
One notes that the two terms in the right-hand side (r.h.s.) of Eq. (B5) are similar in their structure but have opposite signs. This will later show up as a difference of the signs of the two stress tensors, those associated with the "massless" part (the first term in the r.h.s.) and the "massive" part (the second term).

Taking scalar product of both sides of Eq. (B5) with $\mathbf{v}$, one finds after some algebra:
$$\mathbf{v}\cdot\frac{\mathbf{j}\times\mathbf{B}}{c} = -\frac{1}{4\pi}[\mathbf{v}\times\mathbf{B}]\cdot\nabla\times\mathbf{B} + \frac{1}{4\pi\lambdabar^2}[\mathbf{v}\times\mathbf{A}]\cdot\nabla\times\mathbf{A} =$$
$$-\frac{1}{4\pi}\nabla\cdot([\mathbf{v}\times\mathbf{B}]\times\mathbf{B}) + \frac{1}{4\pi\lambdabar^2}\nabla\cdot([\mathbf{v}\times\mathbf{A}]\times\mathbf{A}) -$$
$$\frac{1}{4\pi}\mathbf{B}\cdot(\nabla\times[\mathbf{v}\times\mathbf{B}]) - \frac{1}{4\pi\lambdabar^2}\mathbf{A}\cdot(\nabla\times[\mathbf{v}\times\mathbf{A}]). \tag{B6}$$
Equation (B1) after substituting into it Eq. (4) yields:
$$\nabla\times\left\{\frac{\partial\mathbf{A}}{\partial t} - \mathbf{v}\times\nabla\times\mathbf{A}\right\} = 0, \tag{B7}$$
or
$$\frac{\partial\mathbf{A}}{\partial t} - \mathbf{v}\times\nabla\times\mathbf{A} = \nabla\psi, \tag{B8}$$
where $\psi$ is some scalar function. Now substituting expressions for the vector products from Eqs. (B1) and (B8) into the last line of Eq. (B6), one finds
$$\mathbf{v}\cdot\frac{\mathbf{j}\times\mathbf{B}}{c} = -\frac{1}{4\pi}\nabla\cdot([\mathbf{v}\times\mathbf{B}]\times\mathbf{B}) + \frac{1}{4\pi\lambdabar^2}\nabla\cdot(\psi\mathbf{A}) - \frac{\partial}{\partial t}\left(\frac{B^2}{8\pi}\right) - \frac{\partial}{\partial t}\left(\frac{A^2}{8\pi\lambdabar^2}\right). \tag{B9}$$
The right-hand sides of expressions (B2), (B5) and (B9) contain divergence terms and terms containing the time-derivatives. Collecting the latter group on the left-hand side, one obtains:
$$\frac{\partial}{\partial t}\left(\frac{\rho v^2}{2} + \varepsilon + \frac{B^2}{8\pi} + \frac{A^2}{8\pi\lambdabar^2}\right) = -\nabla\cdot\mathbf{S}, \tag{B10}$$
where the left-hand side can be interpreted as time derivative of energy density, whereas the right-hand side represents divergence of energy flux. For an isolated system, by performing an integration over a volume that encloses this system, one eliminates the divergence terms, and arrives at the energy conservation equation for the Maxwell-Proca MHD:

$$\int_v \left( \frac{\rho v^2}{2} + \varepsilon + \frac{B^2}{8\pi} + \frac{A^2}{8\pi \lambdabar^2} \right) dV = const \ . \tag{B11}$$

The volumetric energy density of the magnetic field, with the Proca term included, is

$$W = \frac{B^2}{8\pi} + \frac{A^2}{8\pi \lambdabar^2} \equiv W_0 + W_1 \tag{B12}$$

In a similar fashion one can derive a virial integral (see, for example, Ryutov, 2008) for the MHD set of equations. This is obtained by taking scalar product of both sides of Eq. (10) with radius-vector *r* (an independent variable) and separating the terms with divergence. Integrating over the volume occupied by the isolated system, one then obtains the virial integral for the Proca MHD:

$$\frac{d}{dt} \int_V \rho v_\alpha x_\alpha dV - \int_V (\rho v^2 + 3p + \frac{B^2}{8\pi} - \frac{A^2}{8\pi \lambdabar^2}) dV + \frac{1}{8\pi G} \int_V g^2 dV = 0 \ . \tag{B13}$$

Here $v_\alpha$ and $x_\alpha$ are the components of the vectors **v** and **r**, respectively, and $G$ is the gravitational constant. [Here and in Appendix C we use Cartesian coordinates so that there is no need to distinguish between the co- and contravariant components of the tensors.]

If the system performs a finite motion, then the time-average of the first term containing time derivative vanishes and one obtains a necessary condition for the system to perform a finite motion: the average of the sum of the rest of the terms must be zero. What is interesting here, is that the Proca term in the magnetic contribution has the same sign as the gravitational pull and thereby adds to the gravitational confinement of the system.

One should remember that the interstellar gas is not an isolated system and neither Eq. (B.11), nor Eq. (B.13) can be applied to it, although these equations can be useful for the qualitative analysis of the role of different effects in the dynamics of this system.

## APPENDIX C AVERAGING THE STRESSES FOR THE RANDOM MAGNETIC FIELD

Now we consider the reduction of the force term in the MHD momentum equation for the case where the field is random, with the correlation scale $l$ much smaller than the scale $a$ at which average parameters of the random field vary. This term can be represented as:

$$f_\alpha = \frac{1}{c} [j \times B]_\alpha = -\frac{\partial \pi_{\alpha\beta}}{\partial x_\beta} \ , \tag{C1}$$

where

$$\pi_{\alpha\beta} = \frac{1}{8\pi} \left[ B^2 \delta_{\alpha\beta} - 2 B_\alpha B_\beta \right] - \frac{1}{8\pi \lambdabar^2} \left[ A^2 \delta_{\alpha\beta} - 2 A_\alpha A_\beta \right] \tag{C2}$$

is the Maxwell-Proca stress tensor (note a typo in Eq. (5) of Ryutov (2008), where a factor of 2 was missing in front of the second term in the square brackets). Taking average over the volume containing many correlation lengths $l$ and assuming average isotropy on the relevant spatial scale, we find that

$$\langle \pi_{\alpha\beta} \rangle = \frac{\delta_{\alpha\beta}}{24\pi}\left[\langle B^2 \rangle - \frac{\langle A^2 \rangle}{\lambdabar^2}\right], \tag{C3}$$

where angular brackets mean the average. We have used the fact that, for the on-average isotropic fluctuations $\langle B_\alpha B_\beta \rangle = (\delta_{\alpha\beta}/3)\langle B^2 \rangle$. We see that the electromagnetic stresses enter the averaged dynamic equations as an isotropic pressure

$$P = \frac{1}{24\pi}\left[\langle B^2 \rangle - \frac{\langle A^2 \rangle}{\lambdabar^2}\right], \tag{C4}$$

with the Proca term giving negative contribution. In the limit of $l \gg \lambdabar$, using Eq. (41) valid in this limit, one can express the Proca contribution in terms of the current density: $\langle A^2 \rangle / 24\pi\lambdabar^2 = 2\pi\lambdabar^2 \langle j^2 \rangle / 3c^2$.

We will use the subscripts "0" and "1" to distinguish between the zero-photon-mass contribution, $P_0 = \langle B^2 \rangle / 24\pi$, and finite-photon-mass contribution, $P_1 = -\langle A^2 \rangle / 24\pi\lambdabar^2$, to the total magnetic pressure:

$$P = P_0 + P_1. \tag{C5}$$

We use the same convention of the meaning of the subscripts "0" and "1" in equations below.

The random magnetic field can be characterized by the spectral distribution over the wave numbers, defined as the magnetic energy per the volume element $dV$ and $\mathbf{k}$-space element $d^3\mathbf{k}$ (Cf. Landau & Lifshitz, 1971, p. 125)

$$dW_0 = \left(\frac{B^2}{8\pi}\right)_k dV d^3\mathbf{k}. \tag{C6}$$

The total magnetic energy in the volume of the cloud is

$$W_0 = \int dV \int \left(\frac{B^2}{8\pi}\right)_k d^3\mathbf{k}. \tag{C7}$$

The spectral distribution may vary across the volume of the cloud, i.e., the integrand of the inner integral may be a function of position inside the cloud. For the further qualitative analysis this point is not significant.

Given that the vector potential is related to the magnetic field by Eq.(4), one can show that the contribution of the vector-potential to the energy integral is:

$$dW_1 = \frac{1}{\lambdabar^2}\left(\frac{A^2}{8\pi}\right)_k dV d^3\mathbf{k} = \frac{1}{\lambdabar^2 k^2}\left(\frac{B^2}{8\pi}\right)_k dV d^3\mathbf{k}. \tag{C8}$$

The contribution of the vector potential to the energy of the whole volume of the cloud is:

$$W_1 = \int dV \int \frac{1}{\lambdabar^2}\left(\frac{A^2}{8\pi}\right)_k d^3\mathbf{k} = \int dV \int \frac{1}{\lambdabar^2 k^2}\left(\frac{B^2}{8\pi}\right)_k d^3\mathbf{k}. \tag{C9}$$

The characteristic spatial scale of the magnetic field variation, as mentioned above, is some $l \ll a$, i.e., the characteristic wavenumber in the spectral distribution of the magnetic field is $1/l$. One can use this for a more formal *definition* of the characteristic length-scale:

$$l^2 = \frac{\int (1/k^2)(B^2)_k d^3\mathbf{k}}{\int (B^2)_k d^3\mathbf{k}} \ . \tag{C10}$$

Note that for the isotropic fluctuations $d^3\mathbf{k} = 4\pi k^2 dk$, in other words, there is no divergence in the upper integral in Eq. (C10). Neglecting the variation of the spectral function across the cloud, one gets a simple relation between the two contributions to the magnetic energy:

$$W_1 = \frac{l^2}{\lambdabar^2} W_0 \ , \tag{C11}$$

where $l^2$ is defined by Eq. (C10).

Referring to potential cosmological applications, we find a parameter $W+3P$ that shows up in the general relativity equations (e.g., Landau & Lifshitz, 1971). The energy density $W$ is, according to Eq. (C11),

$$W = \left(1 + \frac{l^2}{\lambdabar^2}\right) W_0 \ , \tag{C12}$$

whereas the pressure, according to Eqs. (C4) and (C11), is

$$P = \frac{1}{3}\left(1 - \frac{l^2}{\lambdabar^2}\right) W_0 \ . \tag{C13}$$

This yields the following expression for $W+3P$:

$$W + 3P = 2W_0 = \frac{\langle B^2 \rangle}{4\pi} \tag{C14}$$

One sees that $W+3p$ does not experience an enhancement related to the finite photon mass (non-infinite $\lambdabar$), whereas the energy density and the pressure, taken separately, contain an additional parameter $l/\lambdabar$ that can be large, leading to $p \approx -W/3$.

In conclusion, we can formulate the result of the current Appendix in the following form. We deal with slowly varying massive field with vector potential $\mathbf{A}$, $|\mathbf{A}|/\lambdabar \gg |\nabla \times \mathbf{A}| = |\mathbf{B}|$. Vector $\mathbf{A}$ has random orientation is space. In this situation, equation for the negative pressure $p=-W/3$ immediately follows from Eq. (C2) for electromagnetic stress tensor after averaging over the volume containing several correlation lengths of random $\mathbf{A}$.

## APPENDIX D STRESSES IN THE PRESENCE OF A REGULAR FIELD; STRESSES FOR ANISOTROPIC FLUCTUATIONS

In this Appendix we describe two modifications of our basic model: 1) Accounting for the presence of a regular magnetic field; 2) Evaluating the stress tensor for a random magnetic field in the case of the uniaxial anisotropy of the fluctuations.

If there is a regular magnetic field $\mathbf{B}_{reg}$ overlapped with the random, on average isotropic, field, then one needs to add the contribution of the regular field to the overall stress tensor. As the products of the type $B_{reg\alpha} B_\beta$ yield zero upon averaging over many correlation lengths, the averaged contribution of the superposition of the regular and random parts splits into two parts, regular and random, so that the total stress tensor (C2) becomes

$$\pi_{\alpha\beta} = P\delta_{\alpha\beta} + \pi_{\alpha\beta}^{(reg)}, \tag{D1}$$

where the first term is the contribution of a random field, Eq. (C5), and the second is a stress tensor (C2), in which **B** and **A** are those of the regular field. The stresses produced by the regular field are important if $B_{reg} > (l/l_{reg})B$, where $l_{reg}$ is the length-scale of the regular field, $l_{reg} > l$.

Thus far we focused on the fluctuations that are on average isotropic. In the galactic disk this is not necessarily the case: strong "squashing" of the plasma in the $z$ direction (normal to the disk) creates a situation where uni-axial asymmetry naturally appears. The turbulent fluctuations will have different averages for $\langle A_z^2 \rangle$ and for $\langle A_x^2 \rangle = \langle A_y^2 \rangle$, where $x$ and $y$ denote coordinates in the disk plane. Performing the corresponding averaging in Eq. (C2) and retaining only the contribution of the vector potential (that is by far dominant), we find:

$$\pi_{xx} = -\frac{1}{8\pi\lambdabar^2}\left[\langle A_x^2 + A_y^2 + A_z^2 \rangle - 2\langle A_x^2 \rangle\right] = -\frac{\langle A_z^2 \rangle}{8\pi\lambdabar^2}, \tag{D2}$$

$$\pi_{yy} = -\frac{1}{8\pi\lambdabar^2}\left[\langle A_x^2 + A_y^2 + A_z^2 \rangle - 2\langle A_y^2 \rangle\right] = -\frac{\langle A_z^2 \rangle}{8\pi\lambdabar^2}, \tag{D3}$$

$$\pi_{zz} = -\frac{1}{8\pi\lambdabar^2}\left[\langle A_x^2 + A_y^2 + A_z^2 \rangle - 2\langle A_z^2 \rangle\right] = -\frac{\langle A_x^2 + A_y^2 \rangle - \langle A_z^2 \rangle}{8\pi\lambdabar^2}. \tag{D4}$$

Since we assume a uniaxial asymmetry, we have

$$\pi_{xx} = \pi_{yy} \equiv P_d, \tag{D5}$$

where $P_d$ is the pressure acting in the disk plane. It enters the problem of the disk radial equilibrium in the same way as the pressure $P$, Eq. (C4). The pressure

$$\pi_{zz} \equiv P_n \tag{D6}$$

acting in the normal direction, depends on the degree of the anisotropy; $|P_n|$ can be significantly smaller than $|P_d|$.

## APPENDIX E NOTATIONS

*Where possible, we indicate the number of equation in which the corresponding quantity was first introduced*

| | | | |
|---|---|---|---|
| **A** | vector potential (2) | $a_{obs}$ | observed value of centripetal acceleration |
| $a$ | acceleration | | |
| | | $a_R$ | centripetal acceleration (21) |
| **B** | magnetic field (3) | $n_c$ | particle density in molecular cloud |
| **B**$_{ext}$ | external magnetic field (28) | $p$ | gas pressure (10) |
| $c$ | speed of light (1) | $P$ | total magnetic pressure (16) |
| **E** | electric field (3) | $P_0$ | part of $P$ independent of $m_{ph}$ (17) |
| **F**$_{ext}$ | external force (29) | $P_1$ | part of $P$ proportional to $m_{ph}^2$ (17) |
| $f_R$ | centripetal force per unit volume (20) | $R$ | distance to the galaxy center (14) |
| $H$ | thickness of galactic disk (14) | $r_c$ | characteristic size of a cloud |
| $G$ | gravitational constant (B13) | $r_s$ | star radius |
| **g** | gravitational acceleration (10) | $T$ | temperature (19) |
| $\hbar$ | Planck constant (1) | $\hat{T}$ | temporal scale (A8) |
| **j** | current density (5) | $t$ | time |
| **j**$_s$ | current density in stellar material (28) | $u$ | relative velocity of electrons and ions (45) |
| $l$ | correlation length of random magnetic field (14) | **v** | velocity |
| $l_{reg}$ | length-scale of regular magnetic field | v$_A$ | Alfven velocity (47) |
| | | **v**$_{rel}$ | relative velocity of the cloud and the ambient medium |
| $L$ | spatial scale (general) (A8) | | |
| $m_e$ | electron mass | v$_{Te}$ | electron thermal velocity (45) |
| $m_p$ | proton mass | $W$ | energy density of magn. field (C12) |
| $m_{ph}$ | photon mass (1) | $W_0$ | part of $W$ independent of $m_{ph}$ (C7) |
| $M_c$ | mass of a molecular cloud | | |
| $M_s$ | mass of a star (33) | $W_1$ | part of $W$ proportional to $m_{ph}^2$ (C11) |
| $n$ | particle density | $w$ | enthalpy density of gas (B4) |
26

| | | | |
|---|---|---|---|
| $\alpha$ | fraction of centripetal force provided by gravity force (27) | $\mu$ | magnetic moment of a star (29) |
| | | $\Omega$ | angular rotation frequency (19) |
| | | $\rho$ | mass density (9) |
| $\varepsilon$ | internal energy density of gas (12) | $\tau$ | life-time of a compact object |
| $\varphi$ | electrostatic potential (A4) | $\tau_{\rho\varepsilon\sigma}$ | resistive dissipation time of random magnetic field (43) |
| $\eta$ | electrical resistivity (7) | $\tau_{s\text{-}d}$ | frictional slowing-down time of a molecular cloud |
| $\lambdabar$ | photon Compton length (1) | | |

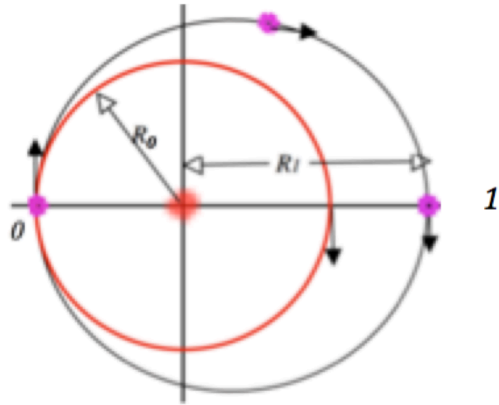

Fig. 1. Transition from rotating gas to a slowly evolving star. The gas rotates along circular trajectory of a radius $R_0$ (red circle) with the center (red fuzzy dot) common for the gravity force and Proca stresses. The star (protostar) is formed quickly at point 1 and after that experiences only gravitational force. The magenta points show consecutive positions of the star as it continues along the black ellipse whose focus is at the center. The largest distance between the gravitating mass and the origin is $R_1$ which is related to $R_0$ by Eq. (37). The figure corresponds to a modest Proca force ($\alpha = 0.2$).